\documentclass[10pt,,notitlepage,superscriptaddress,preprintnumbers,amsmath,amssymb,nofootinbib,aps,prd]{revtex4-1}
\usepackage{amsmath,amssymb,amsfonts}
\usepackage{bm}
\usepackage{graphicx,color}
\usepackage{capt-of}
\usepackage{multirow}
\usepackage{dcolumn}
\usepackage{appendix}
\usepackage{ulem}
\allowdisplaybreaks
\usepackage{hyperref}
\usepackage[all]{hypcap}

\begin{document}
	
	\title{Final state of superradiant instability in Einstein--Maxwell--dilaton black hole}
	\author{A. Rahmani}
	\email{arahmani1090@iau.ac.ir}
	\affiliation{Department of Physics, Varamin-Pishva Branch, Islamic Azad University, Tehran, Iran}
	\author{S. Oghbaiee}
	\email{s_oghbaiee@sbu.ac.ir}
	\affiliation{Department of Physics, Shahid Beheshti University, Tehran 1983969411, Iran}

\begin{abstract}
This article examines the instability of charged black holes, using Einstein--Maxwell--dilaton theory and a massless charged scalar field. Our study demonstrates that these black holes become unstable in a certain range of frequencies, which causes scalar “hair” to form outside the event horizon. By using numerical integration on the coupled field equations, we show that there is a phase transition between a purely charged black hole and a hairy black hole configuration. The results show that the hairy phase appears when the temperature is less than $T_C$, the critical temperature, whereas for $T>T_C$, the system stays purely charged. Also, our findings show that the critical temperature increases with the size of the black hole, implying that the onset of the phase transition becomes progressively more difficult for larger horizons. Lastly, we see that the system enters a stable equilibrium at the critical temperature, which is the final state of the superradiant instability. These results provide us with useful information on how nonlinear superradiant events change and how scalarized black holes arise in more comprehensive theories of gravity.
\end{abstract}
\maketitle

\section{Introduction}

Black holes are among the most remarkable predictions of general relativity (GR). They describe regions of spacetime in which the gravitational field is sufficiently strong that no physical signal, including light, can escape from the event horizon. Black holes are not merely theoretical objects; they play a central role in modern astrophysics and cosmology and are associated with a variety of phenomena observed in the universe, including quasars, gamma-ray bursts, and gravitational-wave events. They also provide natural laboratories for studying gravitational dynamics in the strong field regime and for exploring the connections between classical general relativity, quantum field theory, and possible extensions of the Standard Model (SM)~\cite{Berti:2013uda}.

 Among the phenomena associated with black holes, superradiance is of particular interest. It describes the amplification of certain wave modes scattering from rotating or charged black holes and provides an important mechanism for studying the dynamical behavior and stability of black hole spacetimes. Zel'dovich first identified the phenomenon in rotating systems~\cite{zel1972amplification}, and Press and Teukolsky later developed it, introducing the concept of the “\textit{black hole bomb}”~\cite{press1972floating}. Superradiance has since become an important tool for studying strong-field gravitational systems, particularly in theories that include fields beyond the Einstein--Hilbert framework. Superradiant modes can extract energy and angular momentum from a black hole and may therefore trigger instabilities. These effects are also relevant to gravitational-wave astronomy and to the phenomenology of ultralight particles that arise in extensions of the SM~\cite{dias2012hairy,hod2012instability,hod2016superradiant,hod2014kerr,ponglertsakul2017effect}. 
 
When a mechanism confines the amplified modes near the black hole, superradiant instability can occur. Such confinement can be provided, for example, by a mass term that acts as an effective potential barrier or by a reflecting boundary condition, as in the “\textit{black hole bomb}” mechanism~\cite{brito2014superradiant,brito2015superradiance,press1972floating}. Superradiant instabilities therefore provide a possible probe of physics beyond the SM and offer a way to test general relativity in the strong field regime~\cite{brito2020black,yunes2025gravitational}. 

Within high-energy theory, the dilaton naturally arises as a scalar field in the low-energy limit of string theory and plays a role in determining the effective strength of gauge and gravitational couplings~\cite{green1987superstring,polchinski1998string}. Unlike the metric, which determines the geometry of spacetime, the dilaton is a dynamical field that can modify the coupling between gravity and other fields. As a result, dilaton fields can significantly affect the properties of gravitational systems. In black hole solutions, for example, the dilaton can modify the mass, charge, and scalar hair, providing a connection between general relativity, string theory, and supergravity~\cite{gibbons1988black}.

The Einstein--Maxwell--dilaton--scalar (EMDS) theory has attracted considerable interest as an effective low-energy model motivated by string theory and supergravity~\cite{garfinkle1991charged,gibbons1988black}. The theory consists of gravity coupled to a Maxwell field, a dilaton field that modifies the gauge coupling, and an additional charged scalar field. It therefore admits a richer set of black hole configurations, including charged solutions with scalar hair that can provide examples beyond the conventional \textit{no-hair} picture~\cite{hod2012stationary,garcia2025hairless}. The interaction between charged scalar fields and charged dilatonic black holes can enhance superradiant effects, particularly when the system provides a mechanism for confining the scalar field, such as a reflecting boundary or an asymptotically anti-de Sitter (AdS) spacetime~\cite{dolan2007instability,herdeiro2015asymptotically,he2010signature}. Such interactions can lead to spontaneous scalarization or the formation of stationary scalar condensates around black holes~\cite{gubser2005phase,herdeiro2014kerr,silva2018spontaneous,chen2010holography}.

Unlike asymptotically flat spacetimes, asymptotically AdS spacetime possesses a timelike boundary that can reflect propagating fields toward the bulk, allowing perturbations to interact repeatedly with the black hole~\cite{horowitz2000quasinormal}. This phenomenon is especially important for rotating and electrically charged black holes, which can exhibit superradiant instability. For rotating black holes, superradiance amplification occurs when $\omega < m \Omega_H$, while for charged black holes, the corresponding condition is $\omega < q \Phi_H$, where $ \Omega_H $ and $\Phi_H$ denote the angular velocity and electrostatic potential of the horizon, respectively. Under these conditions, the scattered wave can carry more energy than the incident wave~\cite{gubser2008breaking,cardoso2014holographic}. In AdS spacetime, the reflected radiation can return to the black hole and undergo further amplification, potentially leading to a superradiant instability. The resulting perturbation growth can drive the system toward new equilibrium configurations, including black holes with scalar hair~\cite{cardoso2014holographic,hartnoll2018holographic}. These processes are also relevant to the AdS/CFT correspondence, in which gravitational phenomena in the bulk can be related to phase transitions and other phenomena in the dual boundary theory~\cite{maldacena1999large,hartnoll2018holographic}. 

Despite significant progress in studying superradiant instabilities in different gravitational systems~\cite{brito2015superradiance,herdeiro2015asymptotically}, the role of dilatonic couplings and their interactions with charged scalar fields remains underexplored. Previous studies have mainly considered Kerr or Reissner--Nordström backgrounds~\cite{arvanitaki2011exploring,dias2012hairy}, while including a dynamical dilaton introduces additional effects that can influence both the onset and nonlinear development of the instability~\cite{gibbons1988black,garfinkle1991charged}. In particular, the thermodynamic properties and possible phase transitions associated with hairy black hole configurations in Einstein--Maxwell--dilaton (EMD) theories have not yet been fully understood~\cite{gubser2005phase,herdeiro2014kerr,silva2018spontaneous}. Addressing these questions is important for determining whether dilatonic black holes can represent endpoints of superradiant instabilities and for identifying possible observational signatures relevant to gravitational-wave detection and searches for physics beyond the SM~\cite{brito2017black,yunes2025gravitational}. 

This study examines the superradiant instability of charged black holes in the EMD framework with a massless charged scalar field in AdS spacetime. We focus in particular on the critical temperature associated with the onset of the instability and the possible phase transition between the charged and hairy phases. We first construct the EMDS model and derive the corresponding equations of motion. We then study spherically symmetric charged scalar perturbations to determine the conditions for the onset of superradiance. Based on these results, we construct static, spherically symmetric black hole solutions with charged scalar hair and investigate their thermodynamic properties to determine whether they can represent the endpoint of the instability. Finally, we examine the dynamical stability of the hairy black hole solutions by introducing linear perturbations around the corresponding static backgrounds and studying their time evolution. Stable configurations are characterized by decaying perturbation modes, whereas growing modes indicate an instability. Our results provide a detailed picture of the connection between superradiant amplification, the formation of scalar hair, the stability of hairy black holes, and their thermodynamic behavior in the EMDS framework.

\section{ Setup and Field Equations \label{BHa}}

We consider a massless charged scalar field propagating in the framework of EMD gravity. The dynamics of the system are described by the action~\cite{ran,cadoni2010phase,richarte2022scattering}
\begin{equation}\label{action}
    S = \int d^4 x \sqrt{-g} \left[R - 2 \nabla_{\mu} \Phi \nabla^{\mu} \Phi - e^{-2 \Phi} F^2 - V(\Phi) - D_{\mu}^* \Psi^* D^{\mu} \Psi \right],
\end{equation}
where $R$ is the Ricci scalar, $\Phi$ denotes the dilaton field, and $F^2= F^{\mu \nu} F_{\mu \nu}$ with $ F_{\mu \nu} = \nabla_{\mu} A_{\nu} -\nabla_{\nu} A_{\mu}$, represents the Maxwell invariant. The gauge-covariant derivative acting on the charged scalar field $\Psi$ is defined by $D_{\mu}=\nabla_{\mu}-i q A_{\mu}$, where $ A_{\mu}$ is the electromagnetic vector potential and $q$ is the charge of the scalar field. The asterisk (*) denotes complex conjugation. Throughout this work, we use the natural units $8 \pi G = C = 1$ for convenience. The dilaton potential is chosen to ensure that the spacetime is asymptotically AdS and is given by~\cite{ran}
\begin{equation}\label{potential}
    V(\Phi) = -\frac{4}{L^2} - \frac{1}{L^2}\left[ e^{2 (\Phi - \Phi_0)} + e^{-2 (\Phi - \Phi_0)}\right],
\end{equation}
where $L$ and $\Phi_0$ are the AdS radius and a constant parameter that determines the asymptotic value of the dilaton field, respectively.

The equations of motion are obtained by varying the action~(\ref{action}) with respect to the metric, electromagnetic potential, charged scalar field, and dilaton field. The gravitational field equation takes the form
\begin{equation}\label{eom}
    R_{\mu \nu} - \frac{1}{2} g_{\mu \nu} R = T_{\mu \nu}^F + T_{\mu \nu}^{\Psi} + T_{\mu \nu}^{\Phi},
\end{equation}
where the total stress-energy tensor receives contributions from the Maxwell field, the charged scalar field, and the dilaton sector. The Maxwell equation is given by
\begin{equation}\label{eom-f}
    \nabla_{\mu} (e^{-2 \Phi} F^{\mu \nu}) = -\frac{q}{4} \left[ -2 q A^{\nu} \Psi^* \Psi + i \left( \Psi \nabla^{\nu} \Psi^* - \Psi^* \nabla^{\nu} \Psi \right)\right],
\end{equation}
and the charged scalar field satisfies
\begin{equation}\label{eom-psi}
    D_{\mu} D^{\nu} \Psi =0.
\end{equation}
The equation of motion for the dilaton field is written as
\begin{equation}\label{eom-phi}
    \nabla_{\mu} \nabla^{\mu} \Phi + \frac{1}{2 L^2} \left( e^{2 (\Phi - \Phi_0)} - e^{-2 (\Phi - \Phi_0)} \right) = e^{-2 \Phi} (\nabla_{\mu} A_{\nu} - \nabla_{\nu} A_{\mu})  \nabla^{\nu} A^{\mu}.
\end{equation}

The corresponding components of the stress-energy tensor are expressed as
\begin{equation}\label{t-f}
    T_{\mu \nu}^F = e^{-2 \Phi} \left( -2 \nabla_{\mu} A^{\eta} \nabla_{\nu} A_{\eta}  + 2 \nabla_{\mu} A_{\eta} \nabla^{\eta} A_{\nu} + 2 \nabla_{\nu} A_{\eta} \nabla^{\eta} A_{\mu} -2 \nabla_{\eta} A_{\nu} \nabla^{\eta} A_{\mu} - g_{\mu \nu}  \nabla_{\eta} A_{\kappa} \nabla^{\kappa} A^{\eta} +g_{\mu \nu}  \nabla_{\kappa} A_{\eta} \nabla^{\kappa} A^{\eta} \right),
\end{equation}
\begin{equation}\label{t-psi}
    T_{\mu \nu}^\Psi = \frac{1}{2} \left( D_{\mu}^* \Psi^* D_{\nu} \Psi + D_{\mu} \Psi D_{\nu}^* \Psi^* - g_{\mu \nu}  D^{* \eta} \Psi^* D_{\eta} \Psi \right),
\end{equation}
\begin{equation}\label{t-phi}
    T_{\mu \nu}^\Phi = - 2 \nabla_{\mu} \Phi \nabla_{\nu} \Phi + g_{\mu \nu} \nabla_{\eta} \Phi \nabla^{\eta} \Phi - \frac{1}{ 2 L^2} g_{\mu \nu} \left( 4 + e^{2 (\Phi - \Phi_0)} + e^{-2 (\Phi - \Phi_0)} \right).
\end{equation}
It is important to note that in the limit $\Phi \rightarrow 0$, the dilatonic sector is switched off and the equations of motion reduce to those of the EMS field theory. Therefore, the present EMDS framework can be regarded as a natural extension of the EMS model, in which the dilaton field introduces an additional degree of freedom coupled to the electromagnetic sector. 
A charged static and spherically symmetric AdS black hole solution is given by~\cite{ran}
\begin{equation}\label{metric}
    ds^2 = - N(r) dt^2 + \frac{dr^2}{N(r)} + r^2 g(r) d\Omega^2,
\end{equation}
where the function $g(r)=1-\frac{Q^2 e^{ -2 \Phi_0}}{r M}$, and the metric function $N(r)$ takes the form~\cite{ran}
\begin{equation}\label{def-n}
    N(r) = 1 - \frac{2 M}{r} + \frac{r^2 g(r)}{L^2}.
\end{equation}
Here, $M$ and $Q$ denote the mass and electric charge of the black hole, respectively. This background provides the charged dilatonic geometry on which the scalar perturbations are studied and builds the starting point for analyzing both the superradiant instability and the formation of hairy black hole configurations.

\subsection{ Superradiant Instability in the EMDS Theory\label{BHa1}}

To examine the conditions for superradiant scattering, we first neglect the backreaction of the charged scalar field on the background geometry. This approximation allows us to study the propagation of scalar perturbations on a fixed charged black hole spacetime. It also helps identify the frequency regime in which energy extraction from the black hole becomes possible. We consider a monochromatic charged scalar perturbation characterized by frequency $\omega$ and azimuthal quantum number $m$ which may be decomposed as~\cite{ra,ra1,ra2,rahmani2020instability} 
\begin{equation}\label{def-Psi}
    \Psi(t,r,\theta,\varphi) = \frac{\psi(r)}{r} Y_{lm}(\theta,\varphi) e^{-i \omega t + i m \varphi},
\end{equation}
where $Y_{lm}(\theta,\varphi)$ denotes the usual spherical harmonic.

By substituting the above perturbation into Eq.~(\ref{eom-psi}) and introducing the tortoise coordinate defined by $\frac{dr_*}{dr}=\frac{1}{Ng}$, the radial equation for the charged scalar field can be written in a Schrödinger-like form as
\begin{equation}\label{schrod1}
    \frac{d^2 \psi}{{dr_*}^2} + V_{eff} \psi = 0,
\end{equation}
where the effective potential is given by
\begin{equation}\label{v-eff}
    V_{eff} = g^2 (\omega - \frac{q Q}{r})^2 - N g^2 \left( \frac{N'}{r} - \frac{N g'}{r g} + \frac{L(L+1)}{r^2 g} \right),
\end{equation}
here, the gauge field is chosen as $A_{\mu}=\left(-\frac{Q}{r},0,0,0\right)$, where $Q$ denotes the black hole charge, $q$ is the charge of the scalar field, and the prime indicates differentiation with respect to the radial coordinate $r$.

Near the event horizon and in the asymptotic region, the scalar field admits the following limiting forms
\begin{eqnarray}
    && \psi \sim T e^{- i \tilde{\omega} r_*} \qquad \qquad \qquad r_* \rightarrow -\infty, \label{transmit}
    \\
    && \psi \sim I e^{- i \omega r_*} + R e^{ i \omega r_*}  \qquad r_* \rightarrow +\infty, \label{reflect}
\end{eqnarray}
where $\tilde{\omega}=g(r_h)\left(\omega-\frac{qQ}{r_h}\right)$, and $r_h$ denotes the radius of the event horizon. In the above expressions, $T$, $I$, and $R$ correspond to the transmission coefficient, the amplitude of the incident wave, and the reflection coefficient, respectively. By evaluating the Wronskian relation between the near-horizon and asymptotic solutions given in Eqs.~(\ref{transmit})  and~(\ref{reflect}), one obtains 
\begin{equation}\label{coef-rel}
    I^2 - \frac{\tilde{\omega}}{\omega} T^2 = R^2.
\end{equation}

Superradiant scattering occurs when the reflected wave carries more energy than the incident wave, namely when the scalar mode is amplified after interacting with the charged black hole background. From Eq.~(\ref{coef-rel}),  this amplification requires $\tilde{\omega}<0$, which implies $R>I$. Therefore, the condition for superradiance is given by this inequality 
\begin{equation}\label{sup-cond}
    \omega < q \Phi_h.
\end{equation}
This inequality shows that the frequency of the charged scalar perturbation must be below the product of the scalar charge and the electrostatic potential at the black hole horizon, $\Phi_h = \frac{Q}{r_h}$. In Ref.~\cite{ran}, it was shown that such a superradiant instability can occur for both small and large black holes. It indicates that the instability is not restricted to a particular range of horizon radii. This result suggests that the mechanism responsible for energy extraction and scalar amplification persists throughout different regions of the black hole parameter space. Motivated by this observation, we consider both small and large black hole configurations to obtain a comprehensive understanding of scalar hair formation and the associated phase structure.

To analyze the trapping mechanism responsible for the instability, it is useful to rewrite the radial equation in an alternative form. By redefining the scalar function as $\psi=\frac{R}{r \sqrt{Ng}}$, Eq.~(\ref{schrod1}) can be transformed into
\begin{equation}\label{schrod2}
    R'' + U R =0,
\end{equation}
where the effective potential $U$ is expressed as
\begin{equation}\label{u-eff}
    U = \frac{1}{N^2} \left( \omega - \frac{q Q}{r} \right)^2 + \frac{1}{ r \sqrt{Ng}} \left( (r \sqrt{Ng})'' - \frac{L(L+1)}{r \sqrt{Ng}} \right).
\end{equation}

The effective potential given in Eq.~(\ref{u-eff}) provides important information about the possible confinement of scalar perturbations outside the event horizon. In particular, the formation of a potential well outside the black hole is essential for the development of superradiant instability. This well allows the amplified scalar modes to remain trapped rather than escaping to infinity. This trapping mechanism is combined with the superradiant amplification condition in Eq.~(\ref{sup-cond}); the scalar field can undergo repeated amplification, leading to the growth of bound states and, eventually, to the formation of scalar hair.

\begin{figure}[!ht]
    \includegraphics[width=8cm,height=5cm]{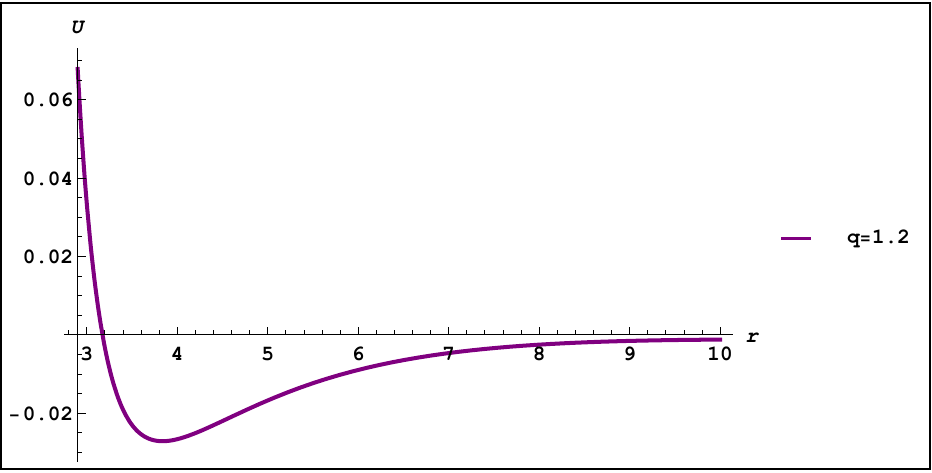}
    \includegraphics[width=8cm,height=5cm]{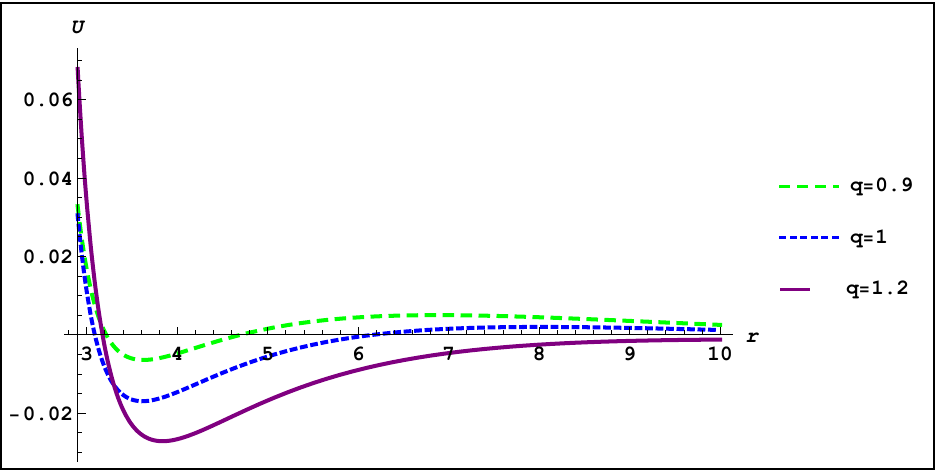}
    \caption{Left: The effective potential as a function of the radial coordinate $r$ for $L=10$, $M=1$, and $Q=0.9$, showing the formation of a potential well outside the event horizon. Right: The effective potential for different values of the scalar field charge.}
    \label{fig1}
\end{figure}

Fig.~\ref{fig1} illustrates the behavior of the effective potential in the EMDS background. The left panel shows the existence of a potential well outside the event horizon. The presence of this well confirms that the geometry can support trapped scalar configurations. The right panel displays the dependence of the potential well on the scalar field charge. The results show that increasing the scalar charge deepens the potential well, thereby strengthening the confinement of scalar perturbations near the black hole. This enhancement of the trapping effect promotes repeated superradiant amplification and increases the likelihood of instability, in agreement with the physical mechanism discussed in Ref.~\cite{hod2013no,basu2016hairy}.

\subsection{Hairy Black Hole Solutions in EMDS Theory}\label{BH1}

Here, we investigate the existence of hairy black hole solutions near the critical temperature. At this temperature, the scalar field begins to condense, and the non-hairy charged black hole may undergo a phase transition toward a scalarized configuration. To construct these solutions, we consider a static and spherically symmetric spacetime ansatz of the form
\begin{equation}\label{br-metric}
    ds^2 = -N(r) h(r) dt^2 + \frac{dr^2}{N(r)}+r^2 g(r) d\Omega^2.
\end{equation}
In this geometry, $h(r)$ translates the backreaction of the matter fields on the spacetime background. In particular, the function $h(r)$ plays an essential role in characterizing the deviation of the hairy black hole solution from the corresponding charged black hole configuration. For the electromagnetic sector, we suppose a purely electric gauge potential $A_{\mu} =(A(r),0,0,0)$, and the scalar field, both depend only on the radial coordinate $r$. This approach preserves the spherical symmetry of the system. It retains the essential physical effects of the charged scalar field, the Maxwell field, and the dilaton coupling as well.

Substituting the metric ansatz~(\ref{br-metric}) together with the matter field configurations into Eqs.~(\ref{eom}--\ref{t-psi}), we obtain a set of coupled nonlinear differential equations. These equations govern the radial evolution of the metric functions, gauge potential, scalar field, and dilaton field and are given by
\begin{equation}\label{cop-eom-1}
    \begin{aligned}
        &(1+ r \Phi') \left(N' h + \frac{N h'}{2} \right) - e^{-2 \Phi} \left(\frac{h e^{2 \Phi_0}}{r}- r {A'}^2 \right)
        \\
        & \quad - \frac{rh}{2 L^2} \left(4 + e^ {2 (\Phi - \Phi_0)} + e^ {-2 (\Phi - \Phi_0)}\right) + \frac{Nh}{r} \left(1+ 4 r \Phi' + 2 r^2 {\Phi'}^2 + r^2 \Phi'' \right)=0,
    \end{aligned}
\end{equation}
\begin{equation}\label{cop-eom-2}
    h' - \frac{r}{(1+ r \Phi')} \left( \frac{q A \Psi}{N} \right)^2 - 4 h \Phi' - r h \frac{{\Psi'}^2 + 2 \Phi''}{(1+ r \Phi')}=0,
\end{equation}
\begin{equation}\label{cop-eom-3}
    A'' + \frac{2 A'}{r} - \frac{h' A'}{2 h} - \frac{q^2 e^{2 \Phi} A \Psi^2}{2 N}=0,
\end{equation}
\begin{equation}\label{cop-eom-4}
    \Psi'' + \left(\frac{2}{r} + \frac{N'}{N} + \frac{h'}{2 h} + 2 \Phi' \right)\Psi' + \left( \frac{q^2 A^2}{N^2 h} \right) \Psi =0,
\end{equation}
\begin{equation}\label{cop-eom-5}
    \Phi'' - \frac{e^{-2 \Phi}}{N} \left( \frac{{A'}^2}{h} + \frac{e^{2 \Phi_0} - e^{2(2 \Phi - \Phi_0)}}{2 L^2} \right) + \left(\frac{2}{r}+\frac{N'}{N} +\frac{h'}{2 h} + 2 \Phi' \right) \Phi'=0.
\end{equation}
The above equations form a highly coupled nonlinear system that generally cannot be solved analytically. Therefore, we use a numerical shooting method to construct the hairy black hole solutions. We integrate the equations from the event horizon, $r=r_h$, toward the reflective boundary. The free parameters at the horizon are fixed to satisfy the required asymptotic boundary conditions. Field regularity at the event horizon imposes important constraints on the near-horizon behavior of the solutions. Specifically, Eqs.~(\ref{cop-eom-2}), (\ref{cop-eom-4}), and Eq.~(\ref{cop-eom-5}) require $A_h=0$, $\Psi'_h=0$, and $\Phi'_h = 0$. These conditions ensure that the gauge, scalar, and dilaton fields remain finite and regular at the horizon. Evaluating Eqs.~(\ref{cop-eom-1}--\ref{cop-eom-5}), at $r=r_h$, yield the corresponding near-horizon boundary conditions as
\begin{equation}\label{horzbc-n}
    N'_h = r_h e^{-2 \Phi_h} \left(\frac{1}{{r_h}^2} - {A'_h}^2 - \frac{1}{2 L} \left( 1+ 4 e^{2 \Phi_h} + e^{4 \Phi_h} \right) \right),
\end{equation}
\begin{equation}\label{horzbc-h}
    h'_h = r_h \left(\frac{q A'_h \Psi_h}{N'_h} \right)^2 + 2 r_h \Phi''_h,
\end{equation}
\begin{equation}\label{horzbc-a}
    A''_h = A'_h \left( - \frac{2}{r_h} + \frac{h'_h}{2} + \frac{q^2 {\Psi_h}^2 e^{2 \Phi_h}}{2 N'_h} \right),
\end{equation}
\begin{equation}\label{horzbc-psi}
    \Psi''_h = -\frac{q^2 {A'_h}^2}{2 {N'_h}^2} \Psi_h,
\end{equation}
\begin{equation}\label{horzbc-phi}
    \Phi''_h = \frac{e^{-2 \Phi_h} A'_h A''_h}{N'_h}.
\end{equation}
Thus, the near-horizon expansion is determined by the horizon data and the regularity conditions given above. For a system to admit well-defined hairy black hole solutions, the horizon values of the gauge field derivative and the dilaton field must satisfy the following constraints
\begin{equation}\label{constraint-a}
    -\frac{1}{\sqrt{2} L} < A'_h < \frac{1}{\sqrt{2} L},
\end{equation}
\begin{equation}\label{constraint-phi}
    \Phi_h = \frac{1}{4} \ln(1 - 2 L^2 {A'_h}^2).
\end{equation}
These relations restrict the allowed parameter space of the solutions and ensure that the resulting configurations remain physically acceptable. Particularly, these conditions are necessary for obtaining regular scalarized black holes that support nontrivial charged scalar hair outside the event horizon.

In addition to the horizon boundary conditions, the scalar field must also satisfy a reflective boundary condition at the asymptotic AdS boundary. Following Ref.~\cite{bosch}, this condition requires the scalar field to vanish at infinity. This boundary condition is essential because the AdS boundary acts as a natural restricting surface. It allows the scalar perturbation to remain trapped outside the black hole. This constraint mechanism plays an important role in the development of superradiant instability and further formation of hairy black hole configurations.

It is also useful to examine the hairy solutions in terms of their Hawking temperatures. This provides direct information about the thermal behavior of the black hole near the critical point. In particular, the formation of scalar hair is closely related to the critical temperature at which the phase transition occurs. For the metric components considered here, the Hawking temperature is given by
\begin{equation}\label{Haw temp}
    T_H = \frac{N'_h \sqrt{h_h}}{4 \pi}.
\end{equation}
This expression shows that this temperature is completely determined by the metric functions near the horizon. Therefore, $T_H$ is a key thermodynamic quantity for studying the onset of scalar condensation, the existence of hairy black hole branches, and the stability of the final configurations generated by the superradiant instability.

\subsection{ Numerical Solutions and Results }

To study the properties of the hairy black hole configurations, we first exploit the scaling symmetry of the system, following the procedure described in Ref.~\cite{oscar}. In particular, the equations of motion are invariant under the transformation
\begin{equation}\label{symmetry}
    (t,r,\theta,\varphi) \rightarrow (\lambda t,\lambda r,\theta,\varphi),\qquad
    (N,h,A,\Phi,\Psi) \rightarrow (N,h,A,\Phi,\Psi),\qquad
    (q,L,r_h) \rightarrow \left(\frac{q}{\lambda}, \lambda L, \lambda r_h \right).
\end{equation}
This symmetry permits a convenient rescaling of the physical quantities and allows us to fix $L= 1$ and $h_h=1$ without loss of generality. Consequently, all dimensional quantities are expressed in units of the AdS radius $L$, yielding dimensionless field equations suitable for numerical analysis.

\begin{figure}[!ht]
    \includegraphics[width=8cm,height=5cm]{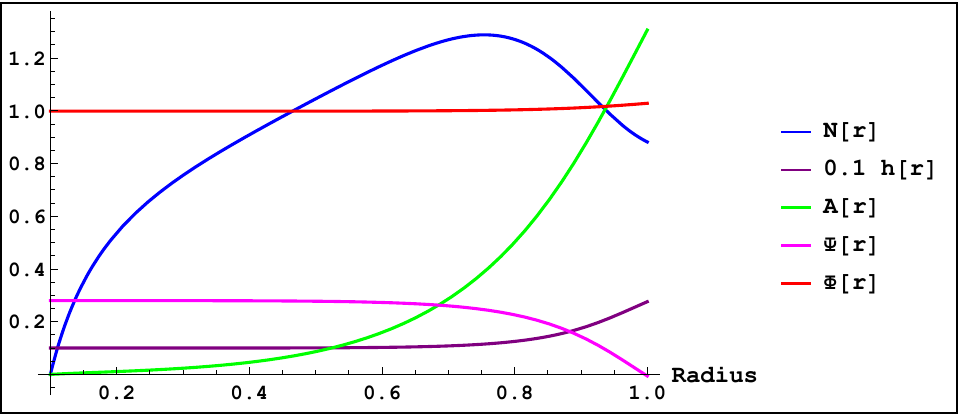}
    \includegraphics[width=8cm,height=5cm]{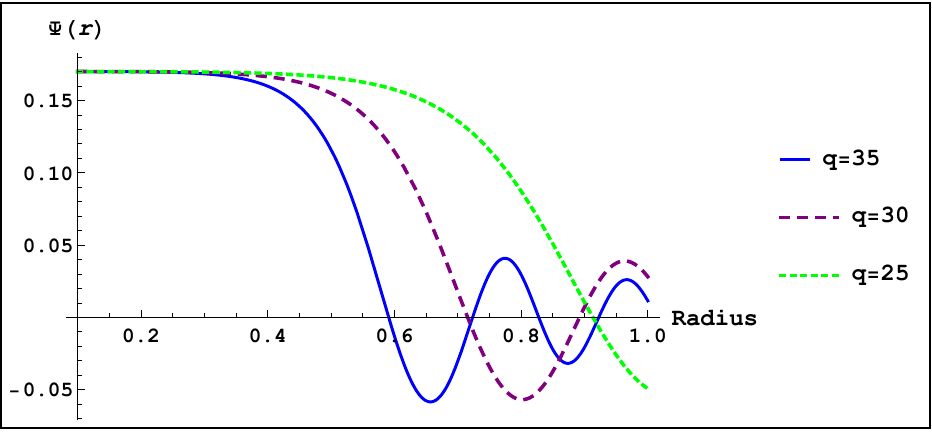}
    \caption{Left: Radial profiles of the field variables for a hairy black hole configuration with $r_h=0.1$, $A'_h=0.146$, $q=15$, and $\Psi_h=0.28$. Right: Radial behavior of the scalar field $\Psi(r)$ for fixed $A'_h=0.146$, $\Psi_h=0.17$, and $r_h=0.1$, considering different values of the scalar charge $q$.}
    \label{fig2}
\end{figure}

The numerical behavior of the resulting solutions is presented in Fig.~\ref{fig2}. The left panel displays the radial profiles of the metric, gauge, dilaton, and scalar field functions, demonstrating that regular, non-singular hairy black hole solutions exist outside the event horizon within the EMDS framework. The right panel shows how the scalar field profile varies with the scalar charge $q$. As the scalar field charge increases, the scalar field exhibits more apparent oscillatory behavior, with a larger number of nodes. This behavior suggests that higher scalar field charges are associated with stronger instability modes in the black hole background. 

\begin{figure}[!ht]
    \includegraphics[width=8cm,height=5cm]{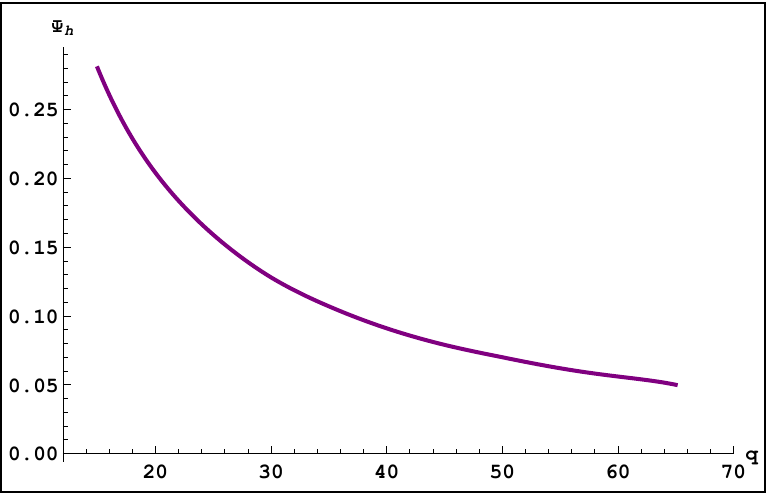}
    \includegraphics[width=8cm,height=5cm]{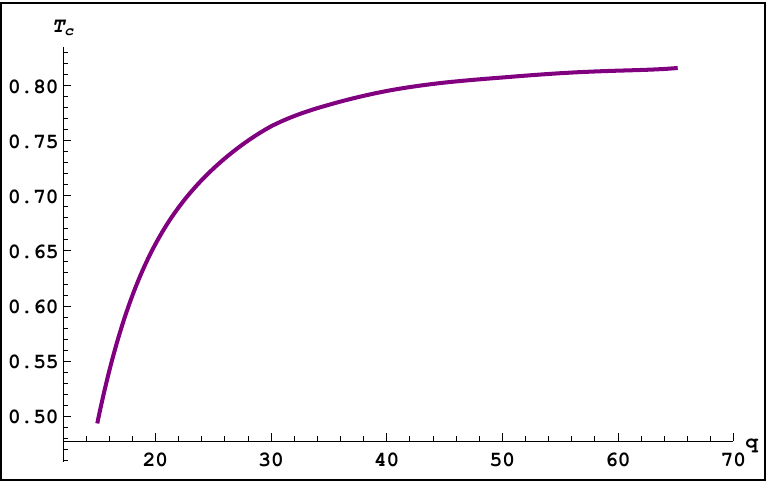}
    \caption{Left: The horizon value of the scalar field $\Psi_h$ as a function of the scalar charge $q$. Right: The critical temperature, $T_C$, as a function of $q$. In both plots, the solutions correspond to configurations with a single node of the scalar field at the reflecting boundary for $A'_h=0.146$ and $r_h=0.1$.}
    \label{fig3}
\end{figure}

In the left panel of Fig.~\ref{fig3}, the horizon value of the scalar field, $\Psi_h$, is examined as a function of the scalar charge, with the scalar field required to have a single node at the reflective boundary. The result shows that $\Psi_h$ decreases with increasing scalar charge $q$, indicating an inverse relationship between the horizon value of the scalar field and its charge. The critical temperature, denoted by $T_C$, is defined as the temperature at which the scalar field vanishes. As shown in the right panel of Fig.~\ref{fig3}, the critical temperature increases rapidly with the scalar charge for $q<30$, whereas its growth becomes considerably slower for $q>30$. This indicates that the scalar charge significantly influences the onset of the hairy phase, especially in the low-charge regime.

\begin{figure}[!ht]
    \includegraphics[width=8cm,height=5cm]{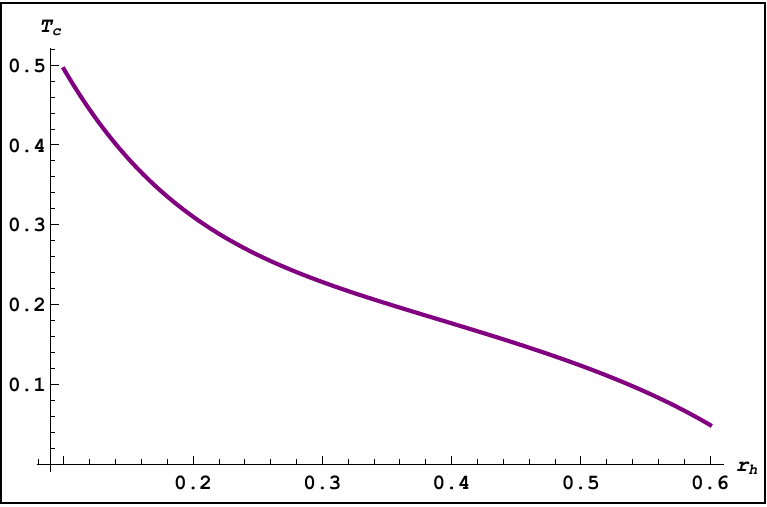}
    \includegraphics[width=8cm,height=5cm]{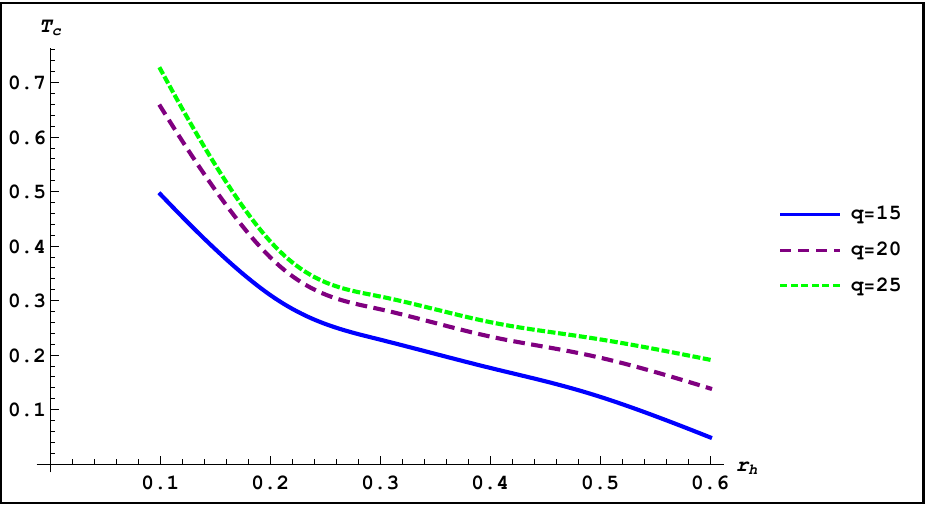}
    \caption{Left: The critical temperature, $T_C$, as a function of the event horizon radius $r_h$ for a fixed scalar charge $q=15$. Right: The dependence of $T_C$ on $r_h$ for different values of the scalar charge $q$ with $A'_h=0.146$.}
    \label{fig4}
\end{figure}

\begin{figure}[!ht]
    \includegraphics[width=8cm,height=5cm]{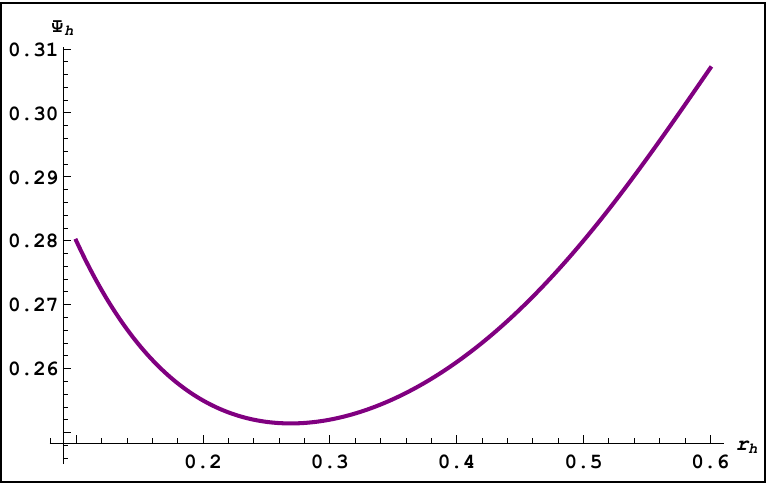}
    \includegraphics[width=8cm,height=5cm]{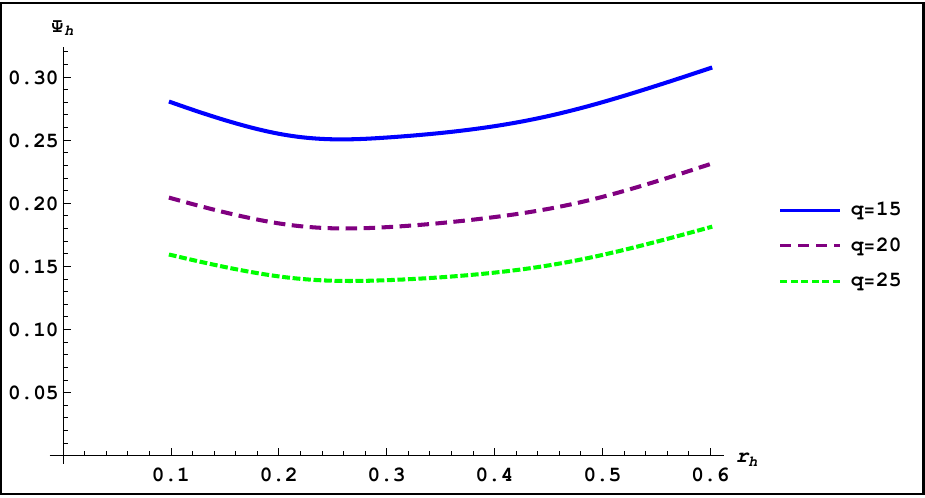}
    \caption{Left: The horizon value of the scalar field, $\Psi_h$, plotted as a function of $r_h$ for a fixed scalar charge $q=15$. Right: The dependence of $\Psi_h$ on $r_h$ for different values of the scalar charge $q$ with $A'_h=0.146$.}
    \label{fig5}
\end{figure}

\begin{figure}[!ht]
    \includegraphics[width=8cm,height=5cm]{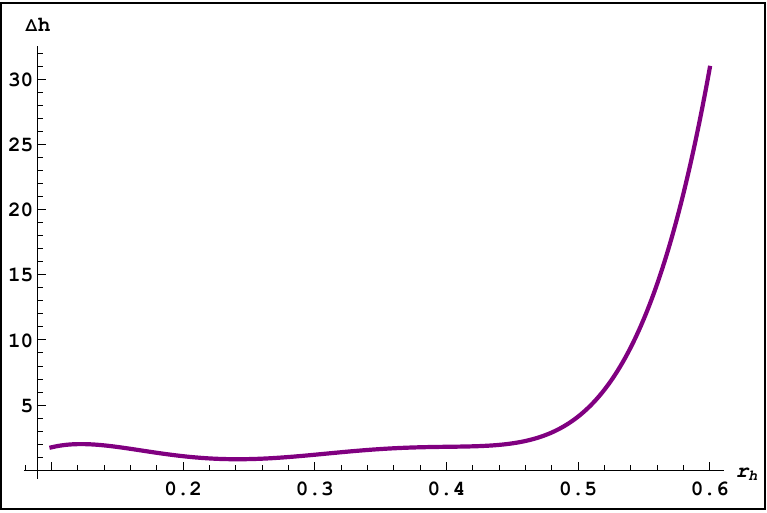}
    \includegraphics[width=8cm,height=5cm]{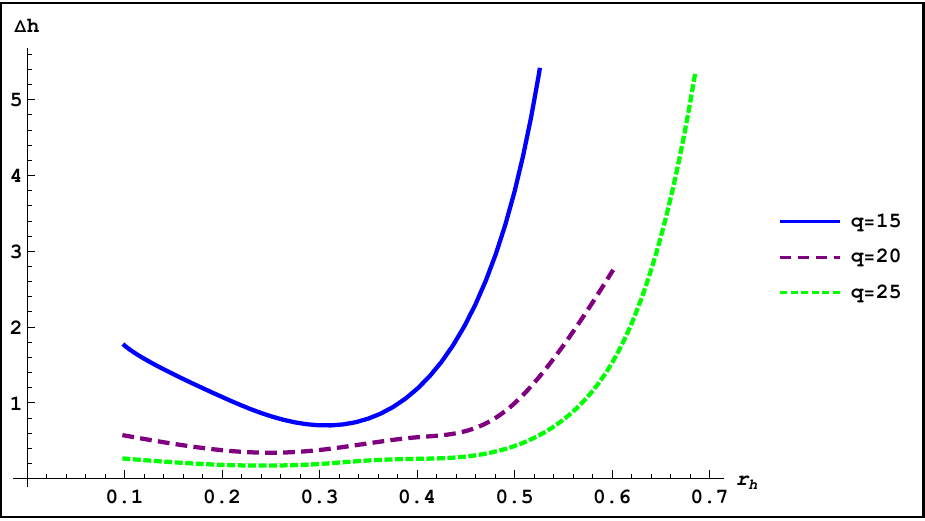}
    \caption{Left: The quantity $\Delta h=h(\infty)-h(r_h)$ as a function of $r_h$ for a fixed scalar charge $q=15$. Right: The behavior of $\Delta h$ as a function of $r_h$ for different values of the scalar charge $q$ with $A'_h=0.146$.}
    \label{fig6}
\end{figure}

\begin{figure}[!ht]
    \includegraphics[width=8cm,height=5cm]{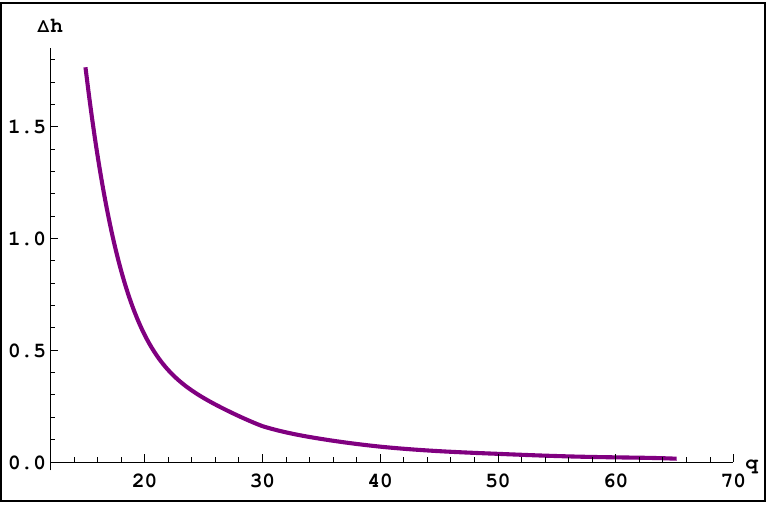}
    \includegraphics[width=8cm,height=5cm]{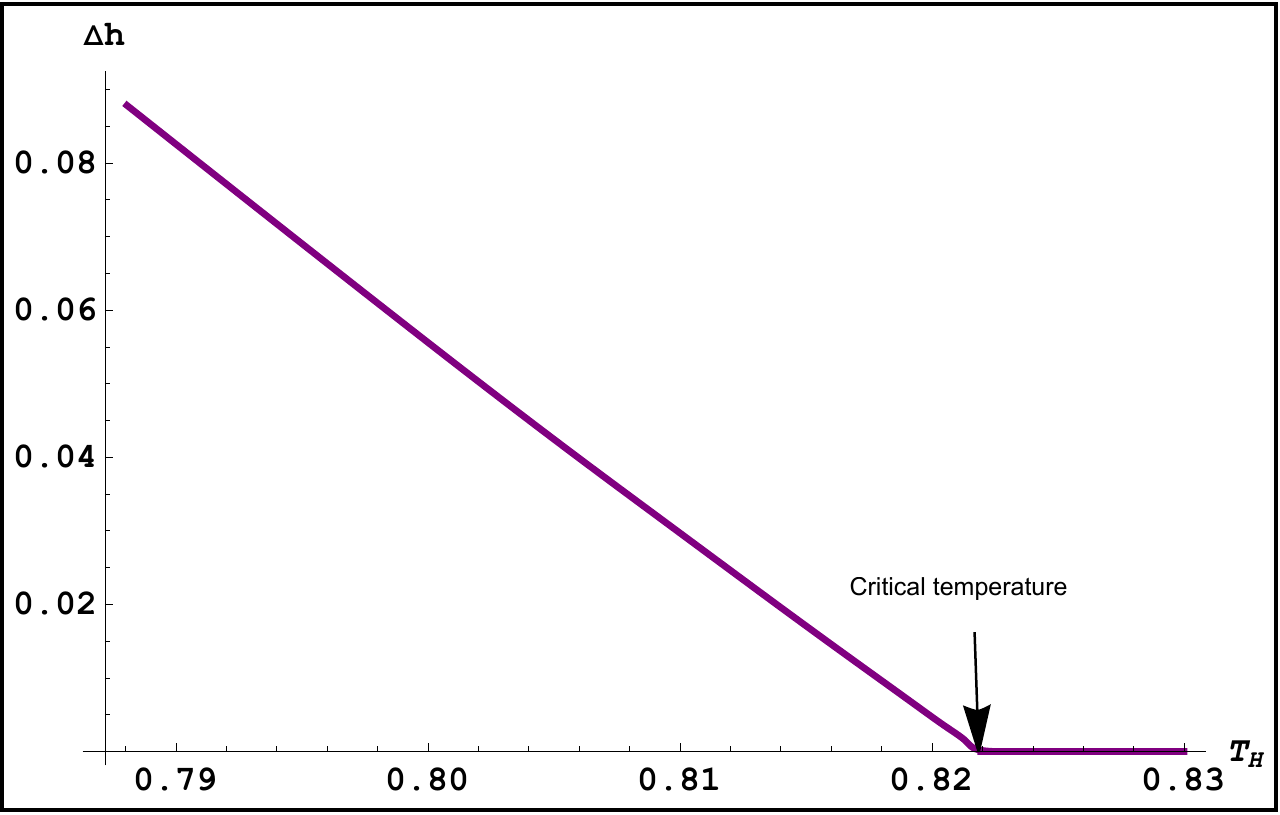}
    \caption{Left: The quantity $\Delta h$ versus the scalar charge $q$. Right: The quantity $\Delta h$ plotted as a function of the Hawking temperature $T_H$. Both plots are drawn at a constant $A'_h=0.146$. The arrow denotes the critical temperature $T_C$ at which the phase transition takes place.}
    \label{fig7}
\end{figure}

\begin{figure}[!ht]
    \includegraphics[width=8cm,height=5cm]{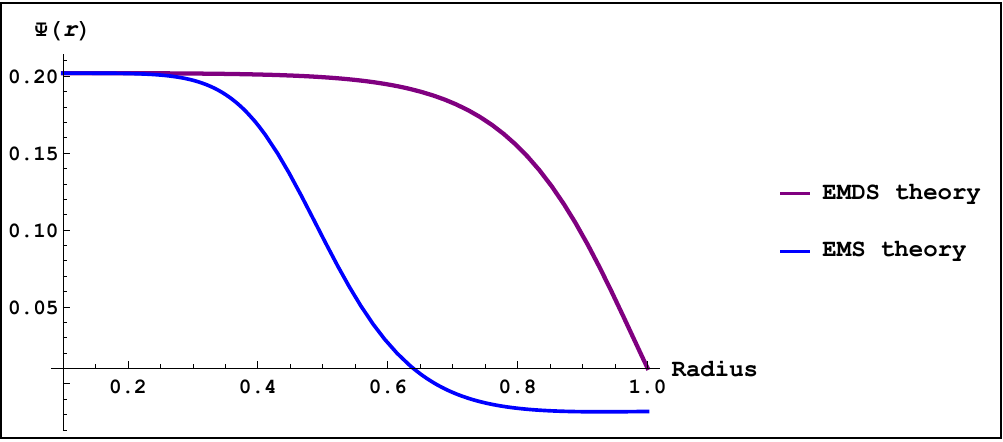}
    \caption{Comparison of the scalar field profiles in the EMDS and EMS theories under similar conditions, with $q = 20$, $A'_h=0.146$, $r_h=0.1$, and $\Psi_h=0.202$.}
    \label{fig8}
\end{figure}

Fig.~\ref{fig4} further illustrates how the critical temperature varies with the event horizon radius and the scalar charge. The left panel shows that $T_C$ decreases as $r_h$ increases, indicating that the smaller black holes experience scalar condensation at higher temperatures. The right panel shows that the critical temperature increases with $q$, meaning that the hairy black hole branch emerges at higher temperatures for larger scalar-gauge coupling. The behavior of the horizon scalar amplitude $\Psi_h$ is shown in Fig.~\ref{fig5}. For small black holes, i.e., $r_h<0.3$, it declines with increasing event horizon radius, indicating an inverse relationship between the scalar field amplitude at the horizon and the black hole size. For larger black holes, with $r_h>0.3$, however, this trend reverses, and $\Psi_h$ grows with the event horizon radius. The right panel also demonstrates that $\Psi_h$ becomes smaller as the scalar charge $q$ grows, confirming that a larger scalar charge suppresses the scalar field amplitude at the horizon. In Fig.~\ref{fig6}, we examine the variation of the metric function through the quantity $\Delta h = h(\infty)–h(r_h) $. The left panel shows that $\Delta h$ varies only slightly for small black holes, whereas it rises sharply in the large black hole regime. This behavior indicates that the matter fields apply a stronger backreaction on the spacetime geometry for sufficiently large horizon radii. The right panel demonstrates that $\Delta h$ is inversely related to the scalar field charge $q$, indicating that a larger scalar charge leads to a smaller variation in the metric function between the horizon and the asymptotic region. The dependence of $\Delta h$ on the scalar charge is displayed more explicitly in the left panel of Fig.~\ref{fig7}. For $q<30$, $\Delta h$ falls rapidly as the scalar field charge increases, while for larger charges $q>30$, it approaches zero. This suggests that the scalar charge significantly reduces the deviation of the hairy black hole configuration from the corresponding background solution in the high charge regime. The right panel of Fig.~\ref{fig7} shows the thermal behavior of the hairy black hole branch. It is evident that regular hairy black hole solutions exist only below the critical temperature, $T<T_C$, whereas no such solutions are present for $T>T_C$. Therefore, the critical temperature marks the phase transition point between the scalarized and non-scalarized black hole phases, as indicated by the arrow in the plot. Finally, Fig.~\ref{fig8} compares the behavior of the scalar field in the EMDS theory with that in the EMS theory. Under similar initial conditions, the scalar field in the EMDS framework exhibits fewer nodes than in the EMS case. This suggests that the presence of the dilaton field modifies the effective scalar dynamics and tends to stabilize the scalar configuration. Hence, black holes in a dilatonic background appear to support more stable scalar field profiles, a conclusion that will be further examined from a thermodynamic perspective in Section~\ref{BHc}.

\section{  The Endpoint of Superradiant Instability \label{BHc}}

In this section, we examine the endpoint of the superradiant instability from a thermodynamic perspective. To determine whether the hairy black hole configurations can represent the thermodynamically preferred endpoint of the instability, their thermodynamic properties must be compared with those of corresponding non-hairy charged black holes. For this purpose, we consider the free energy, which provides a useful diagnostic for studying both thermodynamic stability and possible phase transitions. The free energy is defined as~\cite{basu2016phases,dehghani2019dilaton}
\begin{equation}\label{free ene}
    F=M-TS,
\end{equation}
where $M$ denotes the black hole mass, $T$ is the Hawking temperature, and $S$ is the black hole entropy. The entropy is determined by the horizon area and is given by $S=\frac{A}{4}= \pi {r_h}^2$. Thus, the free energy directly relates the energetic contribution of the black hole mass to the thermal contribution associated with the horizon entropy.

For charged dilatonic black holes, the free energy was obtained in Ref.~\cite{ghaderi} as
\begin{equation}\label{free ene2}
    F=\frac{\pi {r_h}^2 -3 \pi Q^2 (\alpha^2-1)}{4 \pi {r_h}^2},
\end{equation}
while the corresponding black hole mass is given in terms of the entropy by~\cite{ghaderi,nakarachinda2025thermodynamics}
\begin{equation}\label{BH mass}
    M=\frac{S - \pi Q^2(\alpha^2 - 1)}{2 \sqrt{\pi S}}.
\end{equation}
Here, $\alpha$ represents the dilaton coupling constant. Following Ref.~\cite{ghaderi}, we take $\alpha=1$ in the present analysis. This choice specifies the strength of the coupling between the dilaton field and the electromagnetic sector and allows us to compare the thermodynamic behavior of the EMDS solutions with those of the EMS and EM theories under the same physical conditions.

\begin{figure}[!ht]
    \centering
    \includegraphics[width=8cm,height=5cm]{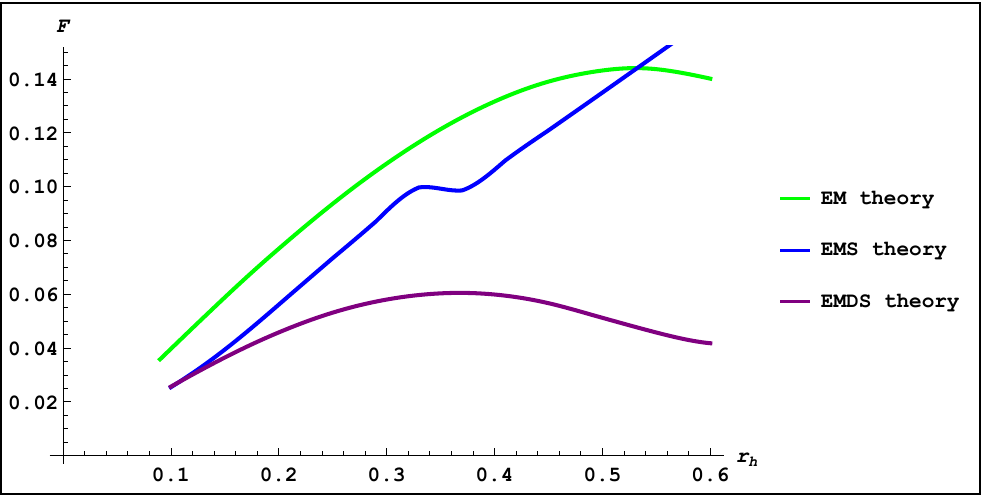}
    \caption{Free energy as a function of the event horizon radius $r_h$ in the EMDS, EMS, and EM theories for $q=25$ and $A'_h=0.3$. The lower free energy of the EMDS configuration indicates that the hairy black hole solution is thermodynamically preferred and represents the more stable phase.}
    \label{fig9}
\end{figure}

We plot the free energy $F$ as a function of the event horizon radius $r_h$ in Fig.~\ref{fig9}. The figure shows that the free energy of the hairy black hole solutions in the EMDS theory is lower than that of the corresponding configurations in the EMS and EM theories. Since the thermodynamically preferred phase is characterized by the lowest free energy, this result indicates that the EMDS hairy black holes are energetically favored. Therefore, from the thermodynamic point of view, the hairy solutions are more stable and can be regarded as the dominant phase of the system.

This behavior supports the interpretation that the endpoint of the superradiant instability in the EMDS framework is a stable hairy black hole rather than the original non-hairy charged black hole. The presence of the dilaton field modifies the thermodynamic structure of the system and lowers the free energy of the scalarized configuration. Consequently, the EMDS hairy black hole branch provides a plausible final equilibrium state after the growth of the charged scalar perturbation.

\section{Stability Analysis}

The results presented in the previous section confirm that the system undergoes a phase transition associated with superradiant instability. In this section, we investigate the endpoint of this instability in the EMDS background and examine whether the final state of a black hole undergoing superradiant instability can be described by a hairy black hole solution at the critical temperature $T_C$.

To establish the existence and stability of the hairy black hole solutions, we first investigate the time evolution of the system. For this purpose, we consider dynamical equations by allowing the field variables to depend on both the radial coordinate and time. For the sake of simplicity, we will omit the explicit dependence of the field variables on $t$ and $r$ in the following equations. By defining $\gamma= N \sqrt{h}$, the corresponding dynamical equations are obtained as follows:
\begin{equation}\label{dyn-nprime}
\begin{aligned}
   \frac{N'}{N}=\frac{r}{2(1+  r \Phi')} \left(\frac{\alpha}{NL^2}-(\frac{2}{r^2}+\frac{4}{r} \Phi'+4 \beta + \psi'{\psi^*}' )- \frac{1}{\gamma^2}(\xi +2 e^{-2\Phi} N {A'}^2 + 2 \frac{\dot{N}}{N}\dot{\Phi}+ \dot{\psi} \dot{\psi^*})\right),
\end{aligned}
\end{equation}
\begin{equation}\label{dyn-hprime}
    \frac{h'}{h}=\frac{r}{\gamma^2(1+  r \Phi')}\left(\xi +\gamma^2 (2\beta + \psi'{\psi^*}')-\frac{\dot{h}}{h}\dot{\Phi}+2(\dot{2\Phi}^2+\ddot{\Phi})+ \dot{\psi} \dot{\psi^*}\right),
\end{equation}
and
\begin{equation}\label{dyn-gammaprime}
    \frac{\gamma'}{\gamma}=\frac{r}{(1+  r \Phi')} \left(\frac{\alpha}{2NL^2}-(\frac{1}{r^2}+\frac{2}{r} \Phi'+ \beta) - \frac{e^{-2\Phi} {A'}^2}{N h} -\frac{1}{\gamma^3} \left( \dot{\gamma} \dot{\Phi} - \gamma (\dot{2\Phi}^2+\ddot{\Phi})\right)\right),
\end{equation}
where the dot denotes the partial derivative with respect to time and the variables $\alpha$, $\beta$, and $\xi$, respectively, are
\begin{equation}\label{alpha}
    \alpha=4+e^{(2\Phi -2)}+(1+\frac{2 L^2}{r^2})e^{-(2\Phi -2)},
\end{equation}
\begin{equation}\label{beta}
    \beta=\frac{2}{r} \Phi'+2 {\phi'}^2+ \Phi'',
\end{equation}
 and
\begin{equation}\label{ksi}
    \xi=q^2 A^2 \psi \psi^* + 2 q A \operatorname{Im}(\psi \dot{\psi^*}).
\end{equation}
The $tr$-component of the Einstein field equations takes the form
\begin{equation}\label{dyn-ndot}
    \frac{\dot{N}}{N}=\frac{r}{(1+  r \Phi')}\left(- \operatorname{Re}(\psi' \dot{\psi^*})- q A \operatorname{Im}
    (\psi {\psi^*}') + ( \frac{N'}{N} + \frac{h'}{h}) \dot{\Phi} -2 ( \frac{\dot{\Phi}}{r} + \dot{\Phi} \Phi' + {\dot{\Phi}}') \right).
\end{equation}
From the Maxwell equation Eq.~(\ref{eom-f}), we obtain two dynamical equations, which can be written as
\begin{equation}\label{dyn-1max}
   \left(\frac{r^2 A'}{\sqrt{h}}\right)'-\frac{r^2 q e^{2 \Phi}}{2 \gamma} \left(q A \psi \psi^* + \operatorname{Im} (\psi \dot{\psi^*})\right)=0,
\end{equation}
and
\begin{equation}\label{dyn-2max}
   \partial_{t}\left(\frac{A'}{\sqrt{h}}\right)-\frac{q \gamma e^{2 \Phi}}{2} \operatorname{Im} (\psi {\psi^*}')=0.
\end{equation}

By defining $\psi = \frac{\Psi}{r}$, the Klein-Gordon equations governing the scalar field can be expressed as
\begin{equation}\label{dyn-scalar}
    \ddot{\Psi}=\gamma^2 \Psi''+\frac{1}{r}( \gamma \gamma'+ 2 \gamma^2 \Phi')(r \Psi'-\Psi )+ \dot{\Psi} \left( \frac{\dot{\gamma}}{\gamma} + 2i q A -2 \dot{\Phi}\right)+\Psi \left(q^2 A^2 - i q A(\frac{\dot{\gamma}}{\gamma}- \frac{\dot{A}}{A}- 2\dot{\Phi} )\right).
\end{equation}
Ultimately, the equation of motion for the dilaton field will be
\begin{equation}\label{dyn-dilaton}
    \ddot{\Phi}= \gamma \gamma' \Phi' + \frac{\dot{\gamma}}{\gamma} \dot{\Phi} - 2 \dot{\Phi}^2 + \gamma^2 \beta - e^{-2 \Phi} N{A'}^2 + \frac{N h}{2 L^2} (e^{(2 \Phi - 2)} - e^{-(2 \Phi - 2)}).
\end{equation}

To examine the stability of the hairy black hole solutions, we introduce linear perturbations around the corresponding static configurations. The field variables are written in the form $N(t,r) = N(r) + \delta N(t,r)$, with analogous expressions for the remaining fields, where $N(r)$ and $\delta N(t,r)$ represent the static background solution and its corresponding perturbation, respectively. Substituting these perturbations into the dynamical equations Eqs.~(\ref{dyn-nprime}--\ref{dyn-dilaton}) and defining $ \delta \psi = \delta u + i \delta \dot{w}$, we obtain a system of three perturbation equations and one constraint, given by
\begin{align}\label{eq1-pert}
        &\delta \ddot{ w}-\gamma ^2 \delta w''
        + \left[\frac{ q^2 A^2 \psi ^2}{r (1+ r \Phi')}+\frac{ q^2 e^{2 \Phi } A \gamma  \sqrt{h}  \psi ^2}{ 2 r^2 A'} - \gamma^2 \left( \frac{\gamma'}{\gamma} + 2 \Phi' \right) \right]\delta w'
        \nonumber 
        \\
        &-\frac{1}{ r} \left[\frac{ q^2 A^2  \left(r+ \psi \psi'+ r^2\Phi' \right)}{(1+ r \Phi')}+\frac{ q^2 e^{2 \Phi }A \gamma  \sqrt{h} \psi' \psi }{ 2 r A'} - \gamma^2 \left( \frac{\gamma'}{\gamma} + 2 \Phi' \right) \right] \delta w 
        -q A \left[2 + \frac{\psi}{(1+ r \Phi')} \left(\frac{\psi}{r}\right)^{\prime} \right]\delta u
        \nonumber 
        \\
        &+\frac{q A \psi }{A'}\delta A'- q \psi \delta A-\frac{2 r q A \psi }{(1+ r \Phi')}\delta \Phi'
        - q A \psi \left[4- \frac{r}{(1+ r \Phi')} \left(\frac{3 h'}{2 h } - \frac{ \gamma' }{\gamma } \right) \right] \delta \Phi = 0,
\end{align}
\begin{align}\label{eq2-pert}
        &\delta \ddot{u} - \gamma^2 \delta u'' - \gamma ^2 \left[\frac{\gamma'}{\gamma } + 2 \Phi' - \frac{r}{(1+ r \Phi')} {\left(\frac{\psi}{r}\right)^{\prime}}^2\right] \delta u' +\left[3 q^2 A^2  + \frac{\gamma^2}{r} \left(\frac{\gamma'}{\gamma} + 2 \Phi' \right) -\frac{q^2 A^2 \psi}{(1+ r\Phi')} \left(\frac{\psi}{r}\right)^{\prime}  \right.
        \nonumber 
        \\
        & \left.  - \frac{ r \gamma^2 } { (1 + r\Phi')} \left(\frac{1}{r} - \frac{h'}{h} + 2 \frac{\gamma'}{\gamma} +  \frac{ \sqrt{h}}{ \gamma (1 + r\Phi')}\left( \frac{\mathcal{E}}{2 r} - \frac{r e^{-2 \Phi } {A'}^2}{h}\right)\right) {\left(\frac{\psi}{r}\right)^{\prime}}^2 \right] \delta u + q \gamma ^2 \left[ 2 A - \psi \left(\frac{A }{(1+ r\Phi')} \right. \right.
        \nonumber 
        \\
        & \left. \left. +\frac{ e^{2 \Phi }\gamma \sqrt{h}  }{2 r A' }\right) \left(\frac{\psi}{r}\right)^{\prime}\right] \delta w'' + q \gamma^2 \left[ 2 A \left( \frac{\gamma'}{\gamma} + 2 \Phi'\right)+ \psi \left( - \frac{A  }{(1+ r \Phi')}\left(\frac{A'}{A} + \frac{h'}{h} - \frac{\gamma'}{\gamma } + 2 \Phi' \right)   \right. \right.
        \nonumber 
        \\
        &\left. \left. +\frac{A \sqrt{h}}{\gamma ( 1+ r \Phi')^2} \left( \frac{\mathcal{E}}{2 r} - \frac{r e^{ - 2 \Phi} {A'}^2}{h}\right) -\frac{ e^{2 \Phi } \gamma \sqrt{h} }{ 2 r A' }\left(\frac{\gamma'}{\gamma} + 2 \Phi'\right) +\frac{ q^2 e^{4 \Phi } A  h \psi^2}{4 r^3 {A'}^2}\right)\left(\frac{\psi}{r}\right)^{\prime}\right] \delta w'
        \nonumber 
        \\
        &+q \left[2 q^2 A^3 - \frac{2 A \gamma^2 }{r } \left( \frac{\gamma'}{\gamma} + 2 \Phi'\right) + \left( - \frac{ q^2 A^3 \psi}{(1+ r \Phi')} + \frac{A \gamma^2 \psi }{r (1+ r \Phi')} \left(  \frac{\gamma'}{\gamma} + 2 \Phi'\right) + \frac{ A \gamma^2 \psi'}{ (1+ r \Phi')} \left( \frac{A'}{A} + \frac{h'}{h} - 2 \frac{\gamma'}{\gamma} \right)\right. \right.
        \nonumber
        \\
        &\left. \left.  - \frac{A \gamma \sqrt{h} \psi'}{( 1+ r \Phi')^2} \left( \frac{\mathcal{E}}{2 r} - \frac{r e^{ - 2 \Phi} {A'}^2}{h}\right) + \frac{e^{2 \Phi} \gamma^3 \sqrt{h} \psi }{2 r^2 A'} \left(\frac{\gamma'}{\gamma} + 2 \Phi'\right) - \frac{ q^2 e^{2 \Phi} A^2 \gamma \sqrt{h} \psi }{2 r A'} - \frac{ q^2 e^{4 \Phi} A \gamma^2 h \psi^2 \psi' }{4 r^3 {A'}^2}
        \right) \left(\frac{\psi}{r}\right)^{\prime}
        \right] \delta w 
        \nonumber
        \\
        &-\frac{r \gamma^2}{A'} \left(\frac{\psi}{r}\right)^{\prime} \delta A'' +\frac{\gamma^2}{ 2 A'} \left[ -4 + r \frac{h'}{h}  + \frac{ q^2 e^{2 \Phi } A \sqrt{h} \psi^2}{ r \gamma A'}\right] \left(\frac{\psi}{r}\right)^{\prime}\delta A' +\frac{2 r^2 \gamma^2}{(1+ r \Phi')} \left(\frac{\psi}{r}\right)^{\prime} \delta \Phi''
        \nonumber 
        \\
        & + r^2 \gamma ^2 \left[\frac{4}{r } + \frac{1}{2 (1+ r \Phi')} \left( \frac{h'}{h} -2 \frac{\gamma'}{\gamma } \right) - \frac{ 2 \sqrt{h}} { \gamma (1+ r \Phi')^2}\left( \frac{\mathcal{E}}{ 2 r} - \frac{r e^{ - 2 \Phi} {A'}^2}{h}\right)\right] \left(\frac{\psi}{r}\right)^{\prime}\delta \Phi' 
        \nonumber 
        \\
        &+\left[4 q^2 A^2 \psi - r^2 \gamma ^2 \left( \frac{1}{r} \left( 3\frac{ h'}{h} - 2 \frac{ \gamma'}{ \gamma}\right) + \frac{1}{2 (1 + r \Phi')} \left( 3 \frac{ h''}{h} - 2 \frac{ \gamma''}{ \gamma} - 5 \frac{ \gamma' h'}{ \gamma h} + 4 \frac{ {\gamma'}^2}{ \gamma^2}\right) \right. \right. 
        \nonumber
        \\
        & \left.\left. -  \frac{ \sqrt{h}} {2 \gamma (1+ r \Phi')^2}\left( \frac{\mathcal{E}}{ 2 r} - \frac{r e^{ - 2 \Phi} {A'}^2}{h}\right) \left( 3 \frac{ h'}{h} -2 \frac{ \gamma'}{ \gamma}\right) \right) \left(\frac{\psi}{r}\right)^{\prime}\right] \delta \Phi = 0,
\end{align}
\begin{align}\label{eq3-pert}
         &\delta \ddot{\Phi} - \gamma^2 \delta \Phi'' - \frac{\gamma^2 }{(1+ 2 r \Phi')} \left[  \frac{2 (1+ r \Phi') (1+ 4 r \Phi')}{r} + \left( \frac{r h' \Phi'}{h} + \frac{\gamma'}{\gamma}\right) - \frac{2 \sqrt{h}}{\gamma (1+ r \Phi')} \left( \frac{\mathcal{H}}{2} + \frac{r e^{- 2 \phi} {A'}^2}{h}\right) \right] \delta \Phi' 
         \nonumber 
         \\
         & - \frac{\gamma \sqrt{h}}{(1+ 2 r \Phi')}  \left[ 2 e^{ 2 \Phi -2} - 4 r \Phi' + \frac{1}{2 (1+ r \Phi')} \left( \frac{\mathcal{H}}{2} + \frac{r e^{- 2 \phi} {A'}^2}{h}\right) \left( 3 \frac{ h'}{h} - 2 \frac{ \gamma'}{\gamma} \right)\right] \delta \Phi
         \nonumber
         \\
         & +\frac{\gamma \sqrt{h}}{r (1+ r \Phi')(1+ 2 r \Phi')} \left[\left( \frac{\mathcal{H}}{2} + \frac{r e^{- 2 \phi} {A'}^2}{h}\right) {\left(\frac{\psi }{r}\right)}^{\prime} \right] \delta u -\frac{q e^{2 \Phi } \gamma ^3 \sqrt{h}  \Phi'  \psi  (1+  r \Phi')}{ r^2 A' (1+ 2 r \Phi')}\delta w''
         \nonumber 
         \\
         & -\frac{ q \gamma^2  \psi (1+  r \Phi') }{r^2 (1+ 2 r \Phi') } \left[ \frac{ A'}{(1+ r \Phi')} + \frac{A \sqrt{h} }{\gamma (1+ r \Phi')^2} \left( \frac{\mathcal{H}}{2} + \frac{r e^{- 2 \phi} {A'}^2}{h}\right) + \frac{ e^{2 \Phi } \gamma \sqrt{h} \Phi' }{A'} \left(  \frac{\gamma'}{\gamma} + 2 \Phi'\right) - \frac{q^2 e^{4 \phi} A h \Phi' \psi^2 }{2 r^2 {A'}^2} \right] \delta w'
         \nonumber 
         \\
         & +\frac{ q \gamma^2  (1+  r \Phi') }{r^2 (1+ 2 r \Phi') } \left[ \frac{ A' \psi'}{(1+ r \Phi')} + \frac{A \sqrt{h} \psi'}{\gamma (1+ r \Phi')^2} \left( \frac{\mathcal{H}}{2} + \frac{r e^{- 2 \phi} {A'}^2}{h}\right) - \frac{q^2 e^{4 \phi} A h \Phi' \psi^2 \psi'}{2 r^2 {A'}^2}  \right.
         \nonumber 
         \\
         & \left. - \frac{e^{2 \Phi} \sqrt{h} \Phi' \psi}{ r \gamma A'} \left( r q^2 A^2 - \gamma^2 \left(  \frac{\gamma'}{\gamma} + 2 \Phi'\right) \right) \right] \delta w -\frac{2 \gamma ^2  \Phi' (1+  r \Phi')}{A' (1+ 2 r \Phi')}\delta A''
         \nonumber 
         \\
         &
         -\frac{\gamma^2  \Phi' (1+  r \Phi') }{ A'(1+ 2 r \Phi') } \left[ \frac{4}{r} - \frac{ h'}{h} - \frac{q^2 e^{2 \Phi } A \sqrt{h}  \psi^2}{r^2 \gamma  A' }\right]\delta A' = 0,
\end{align}
and
\begin{align}\label{eq4-pert}
        &\delta \Phi'' + \left[ \frac{3}{r} + \frac{1}{4} \left(  \frac{h'}{h} - 2\frac{\gamma'}{\gamma} \right) + 7 \frac{ \Phi'}{2} - \frac{\sqrt{h} }{\gamma (1+ r \Phi')} \left( \frac{\mathcal{E}}{2 r} - \frac{r e^{- 2 \phi} {A'}^2}{h}\right)\right] \delta \Phi' - \left[ \frac{q^2 A^2 \psi^2}{2 r^2 \gamma^2 } + \frac{\sqrt{h}}{2 \gamma} \left( 2 + e^{2 \Phi-2} \right) \right.
        \nonumber 
        \\
        & \left. + \left( \frac{(1+ r \Phi')}{4 r} - \frac{\sqrt{h}}{4 \gamma (1+ r \Phi') }  \left( \frac{\mathcal{E}}{2 r} - \frac{r e^{- 2 \phi} {A'}^2}{h}\right) \right) \left( 3\frac{h'}{h} - 2 \frac{\gamma'}{\gamma}\right) +\frac{1}{8} \left( 3 \frac{h''}{h} - 2 \frac{\gamma''}{\gamma} - 5 \frac{\gamma' h'}{\gamma h} + 4 \frac{{\gamma'}^2}{\gamma^2}\right) \right] \delta \Phi
        \nonumber 
        \\
        & + \frac{1}{2 r} \left(\frac{\psi}{r}\right)^{\prime} \delta u'
        - \left[  \frac{q^2 A^2  \psi}{2 r^2 \gamma^2} + \frac{1}{2 r} \left( \frac{1}{r} - \frac{h'}{2h} + \frac{ \gamma'}{\gamma} + \frac{\sqrt{h}}{ \gamma (1+ r \Phi')} \left( \frac{ \mathcal{E}}{2 r} - \frac{r e^{- 2 \phi} {A'}^2}{h}\right) \right) \left(\frac{\psi }{r}\right)^{\prime} \right] \delta u
        \nonumber
        \\
        & - \frac{q A \psi  }{2 r^2 } \delta w'' - \frac{q A \psi }{2 r^2 } \left[ \frac{A'}{A} + \frac{h'}{2 h} + 2 \Phi'- \frac{\sqrt{h}}{ \gamma (1+ r \Phi')} \left( \frac{ \mathcal{E}}{2 r} - \frac{r e^{- 2 \phi} {A'}^2}{h}\right) \right] \delta w'
        \nonumber 
        \\
        & +q \left[ - \frac{q^2 A^3  \psi }{2 r^2 \gamma^2} + \frac{A \psi}{2 r^3} \left( \frac{A'}{A} + \frac{h'}{2 h} + 2 \Phi' - \frac{\sqrt{h}}{ \gamma (1+ r \Phi')} \left( \frac{ \mathcal{E}}{2 r} - \frac{r e^{- 2 \phi} {A'}^2}{h}\right) \right) \right.
        \nonumber
        \\
        & \left. + \frac{A }{ 2 r} \left( \frac{A'}{A} + \frac{h'}{2 h} -\frac{\gamma'}{\gamma}  - \frac{\sqrt{h}}{ \gamma (1+ r \Phi')} \left( \frac{ \mathcal{E}}{2 r} - \frac{r e^{- 2 \phi} {A'}^2}{h}\right) \right) \left(\frac{\psi}{r}\right)^{\prime} \right]\delta w = 0,
\end{align}
where 
\begin{equation}\label{E}
    \mathcal{E} = 4 r^2 + 2 e^{-(2 \Phi -2)} + 2 r^2 \cosh(2 \Phi -2),
\end{equation}
and
\begin{equation}\label{H}
    \mathcal{H} = \mathcal{E} \Phi' - 2 r (1 + r \Phi') \sinh(2 \Phi -2).
\end{equation}
    
Since the perturbation equations are linear, we can integrate them numerically using a shooting method. At the event horizon, we impose ingoing boundary conditions on all perturbation variables,
\begin{equation}\label{boundary}
    \begin{aligned}
        \delta u(t,r) &= \operatorname{Re}\left[ e^{-i \omega(t+r_*)}\, \tilde{u}(r)\right],\\
        \delta w(t,r) &= \operatorname{Re}\left[ e^{-i \omega(t+r_*)}\, \tilde{w}(r)\right],\\
        \delta A(t,r) &= \operatorname{Re}\left[ e^{-i \omega(t+r_*)}\, \tilde{A}(r)\right],\\
        \delta \Phi(t,r) &= \operatorname{Re}\left[ e^{-i \omega(t+r_*)}\, \tilde{\Phi}(r)\right],
    \end{aligned}
\end{equation}
where $r_*$ is the tortoise coordinate defined by $ dr_{*} = \frac{dr}{\gamma}$ and the radial functions are complex. This choice describes waves that fall into the black hole, and it keeps the perturbations regular at the horizon.

To start the integration, we expand each radial function in a Taylor series about the horizon,
\begin{equation}\label{taylor}
    X(r) = \sum_{j=0}^{n} \frac{x_j}{j!} \, (r - r_h)^j,
\end{equation}
and fix the coefficients order by order from the field equations. At leading order, the equations force $\tilde{A}_0$ and $\tilde{\Phi}_0$ to vanish at the horizon, while the horizon values $\tilde{u}_0$  and $\tilde{w}_0$ remain undetermined; all higher-order coefficients in the expansion are fixed in terms of these two constants. They are therefore the only free data of the problem. Since the perturbation equations are linear and homogeneous, however, any solution can be multiplied by an arbitrary complex constant and still be a solution. We use this freedom to set $\tilde{w}_0 = 1$ without loss of generality. The remaining unknowns are then the complex $\tilde{u}_0$ and the complex frequency $\omega$, which are determined together by the two conditions at the AdS boundary, $\tilde{u}(L) = \tilde{w}(L) =0$. This boundary condition ensures that the perturbations remain compatible with the asymptotic structure of the spacetime and allows us to determine the allowed perturbation modes around the hairy black hole solutions.

The resulting eigenfrequencies decide the stability of the hairy black hole solutions. With the time dependence $e^{-i \omega t}$, a mode with $\operatorname{Im}(\omega)<0$ decays in time and corresponds to a stable configuration, whereas $\operatorname{Im}(\omega)>0$ signals a growing mode and hence an instability.

\begin{figure}[!ht]
    \includegraphics[width=8cm,height=5cm]{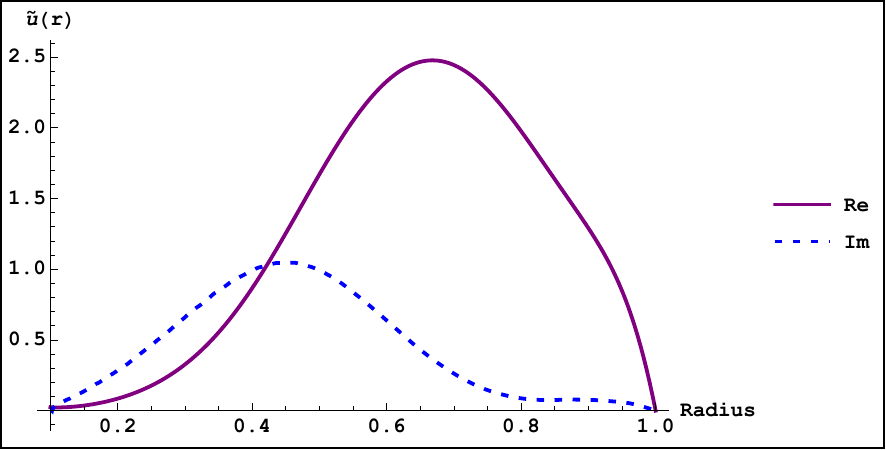}
    \includegraphics[width=8cm,height=5cm]{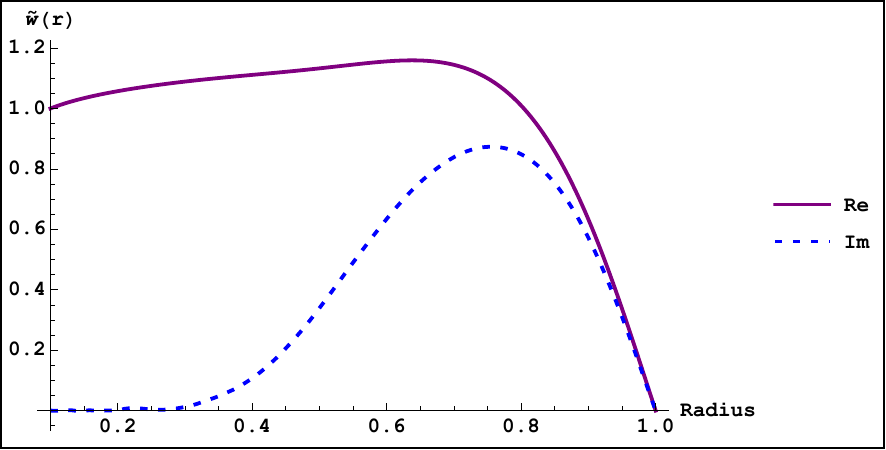}
    \includegraphics[width=8cm,height=5cm]{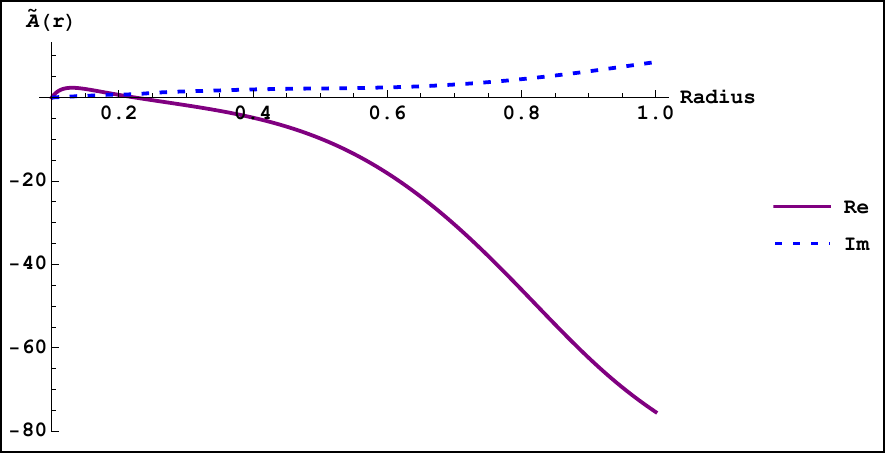}
    \includegraphics[width=8cm,height=5cm]{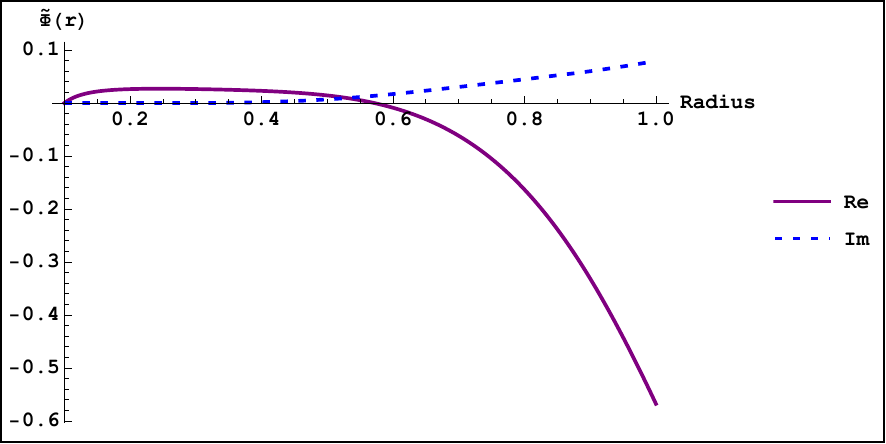}
    \caption{Radial profile of the field variables for the hairy black hole configuration: real (solid) and imaginary (dashed) parts of all modes, for $q=25$, $r_h=0.1$, $A'_h=0.146$, $\psi_h=0.159$, and $\Phi_h=0.998$ with $\tilde{u}_0=0.0241 - 2.26*10^{-14} i$ and $\omega= -1.926*10^{-18} - 0.167 i$. We set the AdS boundary to unity for convenience. Note that the imaginary part has been multiplied by a factor of $10^{10}$ for visibility.}
    \label{fig10}
\end{figure}

Fig.~\ref{fig10} shows the radial functions of $\tilde{u}$, $\tilde{w}$ , $\tilde{A}$, and $\tilde{\Phi} $. In each figure, the real and imaginary parts are plotted together. All functions are regular at the horizon, where $\tilde{A}$ and $\tilde{\Phi}$ vanish as required by the leading-order equations. Both the real and imaginary parts of $\tilde{u}$ and $\tilde{w}$ disappear at $r = L$, confirming that the boundary conditions are satisfied. This shows the black hole's hair is linearly stable against this mode and relaxes monotonically on a time scale $\tau \approx \frac{1}{|\operatorname{Im}(\omega)|} \approx 5.7$.

\begin{figure}[!ht]
    \centering

    \begin{minipage}[c]{0.48\textwidth}
        \centering
        \renewcommand{\arraystretch}{1.89}
        \setlength{\tabcolsep}{25pt}
        \begin{tabular}{|c|c|c|}
            \hline
            $q$ & $\psi_h$ & $\operatorname{Im}(\omega)$ \\ \hline
            15  &  0.280  &  -0.150012 \\ \hline
            20  &  0.204  &  -0.151679 \\ \hline
            25  &  0.159  &  -0.166969 \\ \hline
            30  &  0.128  &  -0.176441 \\ \hline
            35  &  0.107  &  -0.185762 \\ \hline
            40  &  0.091 &  -0.194851 \\ \hline
         \end{tabular}
         \captionof{table}{For $A'_h=0.146$, as $q$ increases, the imaginary part of the frequency decreases almost linearly.}
         \label{tab:1}
    \end{minipage}
    \hfill
    \begin{minipage}[c]{0.48\textwidth}
        \centering
        \includegraphics[width=\linewidth,height=5cm,keepaspectratio]{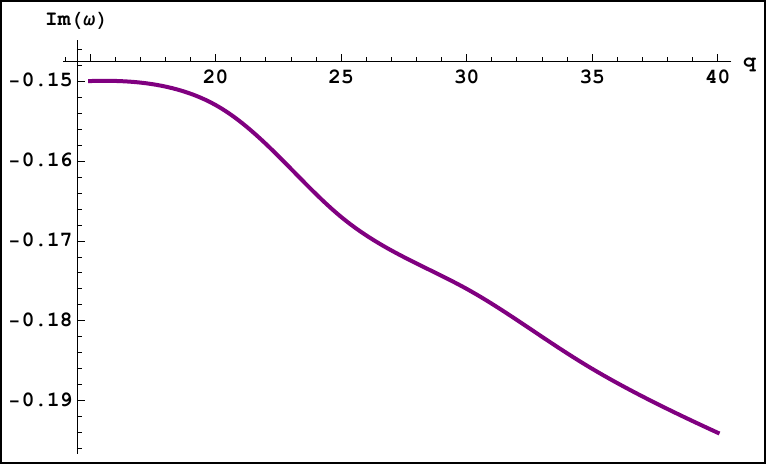}
        \caption{$\operatorname{Im}(\omega)$ as a function of $q$ for $A'_h=0.146$; the data points correspond to the values listed in Table~\ref{tab:1}}
    \label{fig11}
    \end{minipage}
    
\end{figure}

Table~\ref{tab:1} lists the frequencies obtained for several values of $q$, and Fig.~\ref{fig11} displays the corresponding behavior of $\operatorname{Im}(\omega)$. It shows that the amplitude of the perturbation modes decays in time, which is in agreement with the stability of the hairy black hole. Therefore, we may conclude that a black hole with superradiant instability eventually evolves into a hairy black hole.

\section{Discussion and conclusions }\label{DC}

In this work, we demonstrated that superradiant instability can arise in the EMDS theory, with its onset governed by a specific frequency regime, namely
\begin{equation}
    \omega < \frac{q Q}{r_h}. \nonumber
\end{equation}
This condition indicates that charged scalar perturbations can extract energy from the charged black hole background when the wave frequency lies below the corresponding electrostatic threshold. In agreement with the results reported in~\cite{ran}, this instability may arise for both small and large black hole configurations. Furthermore, our analysis revealed an effective potential well outside the event horizon that traps the scalar field. It was shown that increasing the scalar field charge deepens this potential well, thereby enhancing the confinement of the scalar perturbation and strengthening the conditions required for the development of the superradiant instability.

The superradiant instability naturally leads to the formation of hairy black hole solutions, where the hair is described by a charged scalar field localized near the event horizon. We found that the horizon value of the scalar field, $\Psi_h$, exhibits an inverse relation with the scalar field charge $q$. Moreover, the dependence of $\Psi_h$ on the event horizon radius displays distinct behavior in different black hole regimes. For small black holes, $\Psi_h$ decreases with increasing $r_h$, whereas it increases for large black holes. This indicates that, in the large black hole regime, the parameter space supporting hairy configurations becomes broader as the size of the event horizon grows.

We also investigated the thermal properties associated with the emergence of scalar hair. It was shown that the critical temperature $T_C$ is inversely related to the event horizon radius, implying that larger black holes possess lower critical temperatures. This result has important implications for the phase structure of the system. In particular, for large black holes, the lower value of $T_C$ makes the phase transition more difficult to realize, whereas for small black holes, the higher critical temperature facilitates the transition toward the hairy phase. Moreover, the critical temperature varies with the charge of the scalar field according to the relation
\begin{equation}
    T_C \propto \sqrt{q}. \nonumber
\end{equation}
Thus, increasing the scalar charge raises the critical temperature and promotes the formation of scalar hair. An essential quantity in characterizing the phase transition is the metric function $h(r)$, or equivalently, the difference $\Delta h = h(\infty)-h(r_{h})$. This function provides a clear diagnostic for distinguishing the charged black hole phase from the hairy black hole phase. Our results show that regular hairy black hole solutions exist for temperatures below the critical value, $T < T_C$, while no hairy configuration is supported for $T > T_C$. Therefore, the critical temperature marks the transition point at which the scalarized black hole branch disappears, and the system returns to the non-hairy charged black hole configuration.

Finally, from the thermodynamic perspective, our results demonstrate that the hairy black hole solutions are energetically favored over the corresponding non-hairy charged black holes, as they possess lower free energy. We further analyzed the superradiant instability of the charged black hole background and found that its development is consistent with the thermodynamic preference for the hairy configurations. In particular, the instability can drive the system away from the original charged black hole toward a hairy configuration, in agreement with the free energy analysis. These results indicate that the hairy black holes are thermodynamically favored and dynamically preferred and may therefore represent the final equilibrium state of the superradiant instability in EMDS theory. Consequently, our results suggest that the endpoint of the superradiant instability is not the original charged black hole background but rather a stable scalarized black hole endowed with charged scalar hair.

\section{Acknowledgments}

We sincerely thank Dr. Hamid Reza Sepangi for his valuable comments and constructive suggestions, which helped improve this work.

\bibliography{ref}

\end{document}